\documentclass[entropy,article,accept,oneauthors,pdftex,12pt,a4paper]{mdpi}

 \lastpage{2227} \doinum{10.3390/e17042218}
\pubvolume{17(4)} \pubyear{2015}

\history{}

\usepackage{graphics}
\usepackage{amsmath}
\usepackage{graphicx}
\usepackage{amsfonts}
\usepackage{amssymb,soul}

\Title{Cryptographic aspects of quantum reading}

\Author{Gaetana Spedalieri}

\address{Computer Science and York Centre for
Quantum Technologies, University of York, Deramore Lane, York YO10
5GH, United Kingdom}

\corres{}

\abstract{Besides achieving secure communication between two
spatially-separated parties, another important issue in modern
cryptography is related to secure communication in time, i.e., the
possibility to confidentially store information on a memory for
later retrieval. Here we explore this possibility in the setting
of quantum reading, which exploits quantum entanglement to
efficiently read data from a memory whereas classical strategies
(e.g., based on coherent states or their mixtures) cannot retrieve
any information. From this point of view, the technique of quantum
reading can provide a new form of technological security for data
storage.}

\keyword{quantum reading; data storage; security; quantum
entanglement; quantum channel discrimination; quantum fidelity;
Gaussian states}

\begin{document}

\section{Introduction}

Quantum cryptography~\cite{QKDreview,WeeRMP} aims to realize a completely
unbreakable scheme for the distribution of a secret key between two remote
parties, usually called Alice and Bob. Indeed quantum key distribution
(QKD)\ relies its security on one of the the most fundamental physical laws,
the uncertainty principle, which is actively exploited for detecting and
overcoming the presence of a malicious eavesdropper, usually called Eve. In
this scenario, an important role is also played by quantum entanglement~\cite%
{Books}, which can be exploited to make QKD protocols
device-independent, i.e., more robust to practical flaws (e.g., in
the detectors) which may potentially be exploited by Eve. Very
recently, quantum discord~\cite{discordRMP} (see
Ref.~\cite{discordCV} for its computation with Gaussian states)
has also been identified as a useful resource for device-dependent
QKD with trusted noise~\cite{DiscordQKD}, e.g., in scenarios such
as measurement-device independent QKD~\cite{MDI1,MDI2,MDI3,MDI4}.

In this preliminary study, we investigate a different but still
important problem: The confidential storage of information on a
physical device, such as an optical memory. It has been recently
proven that quantum entanglement can provide an advantage in the
readout of classical data from optical memories, especially in the
low-energy regime, i.e., when a few photons are irradiated over
the memory cells. This approach is known as quantum
reading~\cite{Read1} (see also follow-up
papers~\cite{Read2,Read3,Read4,Read5,Read6,Read7,Read8,Read9,Read10,Read11,Read12,Read13}),
a notable application of quantum channel discrimination to a
practical task as the memory readout. From this point of view,
another well-known protocol is quantum illumination, which aims at
improving target
detection~\cite{Lloyd,pirandola,saikat,Shapiro2,Lopaeva,Zhang,ZhangDetection},
and has been recently extended to its most natural domain, the
microwaves~\cite{MicroQI}.

Here we show how the performance advantage given by quantum
reading can be exploited to completely hide classical information
in optical memories. The strategy is to design a photo-degradable
optical memory whose cells have very close reflectivities (each
reflectivity encoding a bit-value). Because of the photodegrable
effects, each cell can only be read with a limited number of
photons. In these low-energy conditions, we find that only
well-tailored quantum sources (in particular, entangled) are able
to discriminate two very close reflectivities and, therefore,
retrieve the information stored in the cell. Specifically, we
derive a simple analytical formula which relates the
reflectivities of the memory cell with the mean number of photons
to be employed by the quantum source.

This approach would provide a layer of technological security to
the stored data, in the sense that only an advanced laboratory
equipped with quantum-correlated sources would be able to read the
information, whereas any other standard optical reader based on
classical states, such as coherent states or even thermal states,
can only extract a negligible number of bits.

The paper is organized as follows. In Sec.~\ref{sec1}, we briefly review the
basic setup of quantum reading and we discuss the performances achievable by
quantum entanglement and classical (coherent) states. Then, in Sec.~\ref%
{sec2} we show how to design memories which are not accessible to classical
methods. Finally, Sec.~\ref{sec3} is for conclusions.

\section{Basic setup for quantum reading}

\label{sec1}

For our purpose we consider the simplest version of quantum reading,
considering only ideal optical memories, i.e., with high reflectivities, and
neglecting decoherence effects (see Ref.~\cite{Read1} for more advanced
models). Each memory cell is assumed to be in one of two hypotheses:
Non-unit reflectivity $r_{0}:=r<1$ (encoding bit-value $0$)\ or unit
reflectivity $r_{1}=1$ (encoding bit-value $1$). Mathematically, this is
equivalent to distinguish between a lossy channel $\mathcal{E}_{r}$ whose
loss parameter is the reflectivity $r<1$ and an identity channel $%
\mathcal{I}$.

In symmetric quantum hypothesis testing, these two hypotheses have
the same cost, so that we aim to optimize the mean error
probability. In other words, we need to minimize
$\bar{p}:=p(1|0)p_{0}+p(0|1)p_{1}$, where $p_{0}$ and $p_{1}$ are
the \textit{a priori} probabilities of the two
hypotheses, while $p(1|0)$ is the probability of a false positive and $%
p(0|1) $ is the probability of a false negative. For simplicity,
we consider here equiprobable hypotheses, i.e., $p_{0}=p_{1}=1/2$,
which means that a bit of information is stored per cell. The
amount of information which is retrieved in the readout process is
therefore given by $ I_{\text{read}}(\bar{p})=1-H(\bar{p})$, where
$H(\bar{p})=-\bar{p}\log _{2}\bar{p}-(1-\bar{p})\log
_{2}(1-\bar{p})$ is the binary formula of the Shannon
entropy~\cite{Cover}.

\subsection{Classical Benchmark}

To distinguish between the two hypotheses Alice exploits an input
source of light (a transmitter) and an output detection scheme (a
receiver). In the classical reading setup, the transmitter
consists of a single bosonic mode, the signal ($S$), which is
prepared in a coherent state $\left\vert \alpha \right\rangle $\
sent to the memory cell. At the output, the receiver is typically
a photodetector counting the number of photons reflected, followed
by a digital processing based on a classical hypothesis test. The
performance of this receiver can be bounded by considering an
optimal quantum measurement, constructed from the Helstrom matrix
$\rho _{0}-\rho _{1}$ of the two possible output states $\rho
_{0}=\left\vert \sqrt{r}\alpha \right\rangle \left\langle
\sqrt{r}\alpha \right\vert $ and $\rho _{1}=\left\vert \alpha
\right\rangle \left\langle \alpha \right\vert $.

The minimum error probability is given by the Helstrom bound~\cite{Helstrom}
which is here very simple to compute since the two states are pure. In fact,
for two arbitrary pure states $\left\vert \varphi _{0}\right\rangle $ and $%
\left\vert \varphi _{1}\right\rangle $, the Helstrom bounds reads%
\begin{equation}
\bar{p}=\frac{1-D(\left\vert \varphi _{0}\right\rangle ,\left\vert \varphi
_{1}\right\rangle )}{2},  \label{pmedio}
\end{equation}%
where the trace distance~\cite{Books} $D$ is determined by the fidelity%
\begin{align}
D(\left\vert \varphi _{0}\right\rangle ,\left\vert \varphi _{1}\right\rangle
)& =\sqrt{1-F(\left\vert \varphi _{0}\right\rangle ,\left\vert \varphi
_{1}\right\rangle )},  \label{traceD} \\
F(\left\vert \varphi _{0}\right\rangle ,\left\vert \varphi _{1}\right\rangle
)& =\left\vert \left\langle \varphi _{0}\right. \left\vert \varphi
_{1}\right\rangle \right\vert ^{2}.  \label{fid}
\end{align}

In our specific case, we have~\cite{WeeRMP}
\begin{equation}
F(\left\vert \sqrt{r}\alpha \right\rangle, \left\vert \alpha
\right\rangle)=\exp \left( -\left\vert \alpha -\sqrt{r}\alpha
\right\vert ^{2}\right) =\exp [-\bar{n}(1-\sqrt{r})^{2}],
\end{equation}%
where $\bar{n}=\left\vert \alpha \right\vert ^{2}$ is the mean number of
photons of the input coherent state. As a result, we achieve the following
Helstrom bound for the coherent state transmitter
\begin{equation}
\bar{p}_{\text{coh}}(\bar{n},r)=\frac{1-\sqrt{1-e^{-\bar{n}(1-\sqrt{r})^{2}}}%
}{2},  \label{CLASStransmitter}
\end{equation}%
which is therefore able to read an average of $I_{\text{read}}^{\text{class}%
}=I_{\text{read}}(\bar{p}_{\text{class}})$ bits per cell.

\subsection{Quantum Transmitter}

In the quantum reading setup, we consider a transmitter composed of two
entangled modes, that we call signal ($S$) and reference ($R$). This is
taken to be an Einstein-Podolsky-Rosen (EPR) state, also known as a two-mode
squeezed vacuum state~\cite{WeeRMP}. An EPR state is a zero-mean pure
Gaussian state $\left\vert \mu\right\rangle _{SR}$ with covariance matrix
(CM)%
\begin{equation}
\mathbf{V}(\mu)=\left(
\begin{array}{cc}
\mu\mathbf{I} & \sqrt{\mu^{2}-1}\mathbf{Z} \\
\sqrt{\mu^{2}-1}\mathbf{Z} & \mu\mathbf{I}%
\end{array}
\right) ,~%
\begin{array}{c}
\mathbf{Z}:=\mathrm{diag}(1,-1), \\
\mathbf{I}:=\mathrm{diag}(1,1),~~%
\end{array}%
\end{equation}
where $\mu\geq1$ quantifies both the mean number of thermal photons in each
mode, given by $\bar{n}=(\mu-1)/2$, and the amount of entanglement between
the signal and reference modes~\cite{WeeRMP}.

The signal mode, with $\bar{n}$ mean photons, is sent to read the memory
cell and its reflection $S^{\prime}$ is combined with the reference mode in
an optimal quantum measurement. Given the state $\rho_{SR}=\left\vert
\mu\right\rangle _{SR}\left\langle \mu\right\vert $ of the input modes $S$
and $R$, we get two possible states
\begin{align}
\sigma_{0} & =(\mathcal{E}_{r}\otimes\mathcal{I})(\rho_{SR}), \\
\sigma_{1} &
=(\mathcal{I}\otimes\mathcal{I})(\rho_{SR})=\rho_{SR},
\end{align}
for the output modes $S^{\prime}$ and $R$ at the receiver. One is
just the input EPR state, while the other state $\sigma_{0}$ is a
mixed Gaussian
state with CM%
\begin{equation}
\mathbf{V}_{0}(\mu,r)=\left(
\begin{array}{cc}
(r\mu+1-r)\mathbf{I} & \sqrt{r(\mu^{2}-1)}\mathbf{Z} \\
\sqrt{r(\mu^{2}-1)}\mathbf{Z} & \mu\mathbf{I}%
\end{array}
\right) .  \label{CM1out}
\end{equation}

The minimum mean error probability is given by the Helstrom bound $\bar{p}_{%
\text{quantum}}=[1-D(\sigma _{0},\sigma _{1})]/2$, where $D(\sigma
_{0},\sigma _{1})$ is the trace distance between $\sigma _{0}$ and $\sigma
_{1}$. The Helstrom bound is difficult to compute when one or both the
output states are mixed. For this reason, we resort to an upper-bound, known
as quantum Chernoff bound (QCB)~\cite{QCB1,QCB2,QCBgauss}. This can be
written as
\begin{equation}
\bar{p}_{\text{quantum}}^{\text{QCB}}:=\frac{C}{2},~~C:=\inf_{s\in
(0,1)}C_{s},
\end{equation}%
where $C_{s}:=\mathrm{Tr}(\sigma _{0}^{s}\sigma _{1}^{1-s})$ is the $s$%
-overlap between the two states. In the specific case where one of
the output states is pure $\sigma _{1}=\left\vert \varphi
\right\rangle \left\langle \varphi \right\vert $, we may write
$C=F$, using the quantum fidelity $F=\left\langle \varphi
\right\vert \sigma _{0}\left\vert \varphi \right\rangle $. For
zero-mean Gaussian states, this fidelity can easily be
computed in terms of their CMs~\cite{Spedalieri13,GaeQHB}. In fact, we have%
\begin{equation}
F=\frac{4}{\sqrt{\det [\mathbf{V}(\mu )+\mathbf{V}_{0}(\mu ,r)]}}=\frac{4}{%
\left[ 1+\mu +\sqrt{r}(1-\mu )\right] ^{2}}=\left( 1+\bar{n}+\bar{n}\sqrt{r}%
\right) ^{-2}.
\end{equation}%
As a result, the mean error probability associated with this quantum
transmitter is upperbounded by the QCB as follows%
\begin{equation}
\bar{p}_{\text{quantum}}\leq \bar{p}_{\text{quantum}}^{\text{QCB}}=\frac{%
\left( 1+\bar{n}+\bar{n}\sqrt{r}\right) ^{-2}}{2}.  \label{EPRtransmitter}
\end{equation}%
Thus, the EPR transmitter is able to read at least $I_{\text{read}}^{\text{%
quant}}=I_{\text{read}}(\bar{p}_{\text{quantum}}^{\text{QCB}})$ bits per
cell.

\section{Data secured by quantum reading}

\label{sec2}

\begin{figure}[tbph]
\vspace{-0.7cm}
\par
\begin{center}
\includegraphics[width=0.6\textwidth] {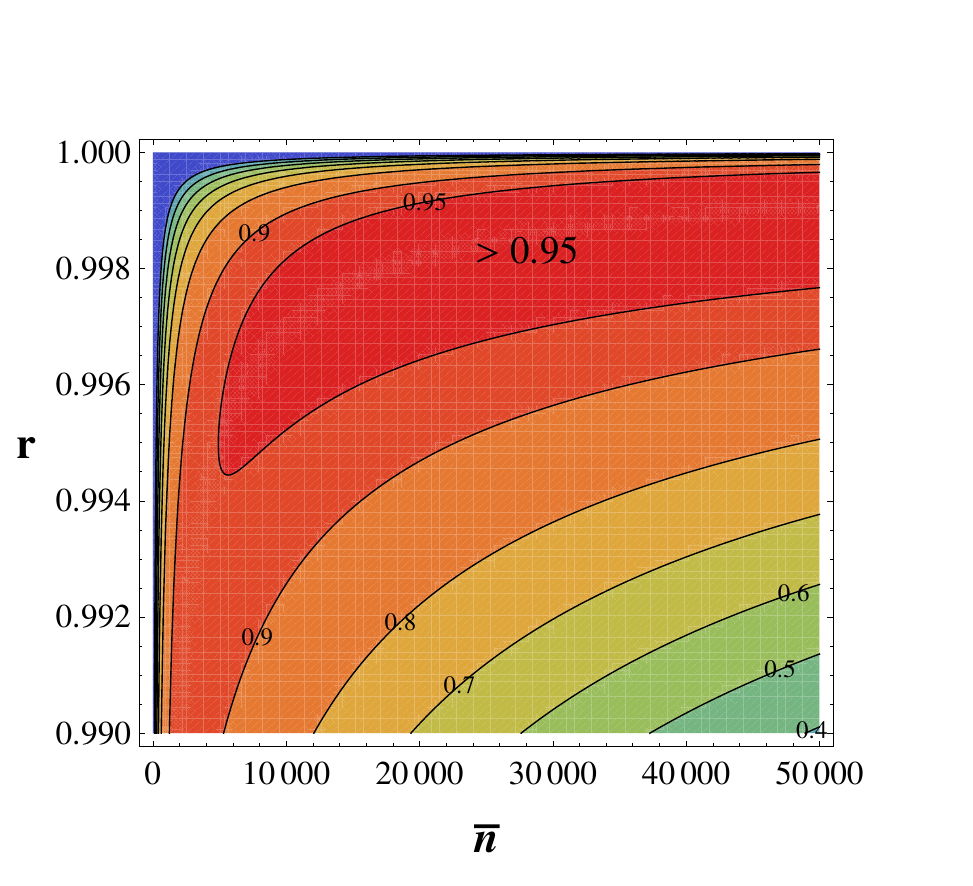}
\end{center}
\par
\vspace{-0.2cm}
\caption{We plot $\Delta (\bar{n},r)$ in the high-reflectivity range $%
0.99\leq r<1$ and wide range of $\bar{n}$ up to $5\times 10^{4}$.
We see how the EPR\ transmitter is superior for $r\simeq 1$, where
$\Delta $ becomes close to 1 bit per cell.} \label{EPRpic}
\end{figure}

We can compare the readout performances of the two transmitters by
considering the information gain $\Delta :=I_{\text{read}}^{\text{quant}}-I_{%
\text{read}}^{\text{class}}$. Its positivity means that quantum
reading outperforms the classical readout strategy. In particular,
for $\Delta \simeq 1$ bit per cell we have that the EPR\
transmitter reads all data, while the classical transmitter is not
able to retrieve any information. Here we aim to exploit this
feature to make the data storage secure in absence of
entanglement (and, more generally, quantum resources). As we can see from Fig.~\ref{EPRpic}, the value of the gain $%
\Delta $ is close to the maximum value of $1$ bit per cell when the memory
cell is characterized by very high reflectivities, i.e., $r\simeq 1$. In
particular, the good region where $\Delta >0.95$ is particularly evident at
low photon numbers, while it tends to shrink towards $r=1$ for increasing
energy.

We now discuss how we can exploit this advantage of quantum
reading for designing a secure classical memory. Let us expand the
information
quantities $I_{\text{read}}^{\text{class}}$ and $I_{\text{read}}^{\text{quant%
}}$ at the leading order in $(1-r)\simeq 0$. We find%
\begin{equation}
I_{\text{read}}^{\text{class}}\simeq \frac{\bar{n}(1-r)^{2}}{\ln 256},~~I_{%
\text{read}}^{\text{quant}}\simeq \frac{\bar{n}^{2}(1-r)^{2}}{\ln 4}.
\label{condF}
\end{equation}%
As we can see, at high reflectivities, there is a different behaviour of
these quantities in the mean number of photons $\bar{n}$. In particular, we
may write
\begin{equation}
I_{\text{read}}^{\text{quant}}\simeq 4\bar{n}I_{\text{read}}^{\text{class}}~.
\end{equation}%
According to Eq.~(\ref{condF}), a non-trivial difference between
$I_{\text{read}}^{\text{class}}$ and
$I_{\text{read}}^{\text{quant}}$ arises by imposing the condition
\begin{equation}
1-r=\bar{n}^{-1}~.\label{condR0}
\end{equation}
Indeed this leads to the following behaviour for large $\bar{n}$%
\begin{equation}
I_{\text{read}}^{\text{class}}\simeq \frac{1}{\bar{n}\ln 256}\rightarrow
0,~~I_{\text{read}}^{\text{quant}}\simeq \frac{\ln \left( \frac{2048}{81}%
\right) -7\ln \left( \frac{9}{7}\right) }{\ln 512}\simeq 0.235.
\end{equation}%
We can see that only quantum reading enables to retrieve non-zero
information from the memory (combining this performance with
suitable error correcting codes would enable us to achieve a
complete readout of the memory). In the following
Fig.~\ref{EPRpic2}, we show the behaviour of
the two information quantities $I_{\text{read}}^{\text{class}}$ and $I_{%
\text{read}}^{\text{quant}}$ in terms of the mean photon number $\bar{n}$\
and assuming the condition of Eq.~(\ref{condR0}).
\begin{figure}[tbph]
\vspace{-0.7cm}
\par
\begin{center}
\includegraphics[width=0.52\textwidth] {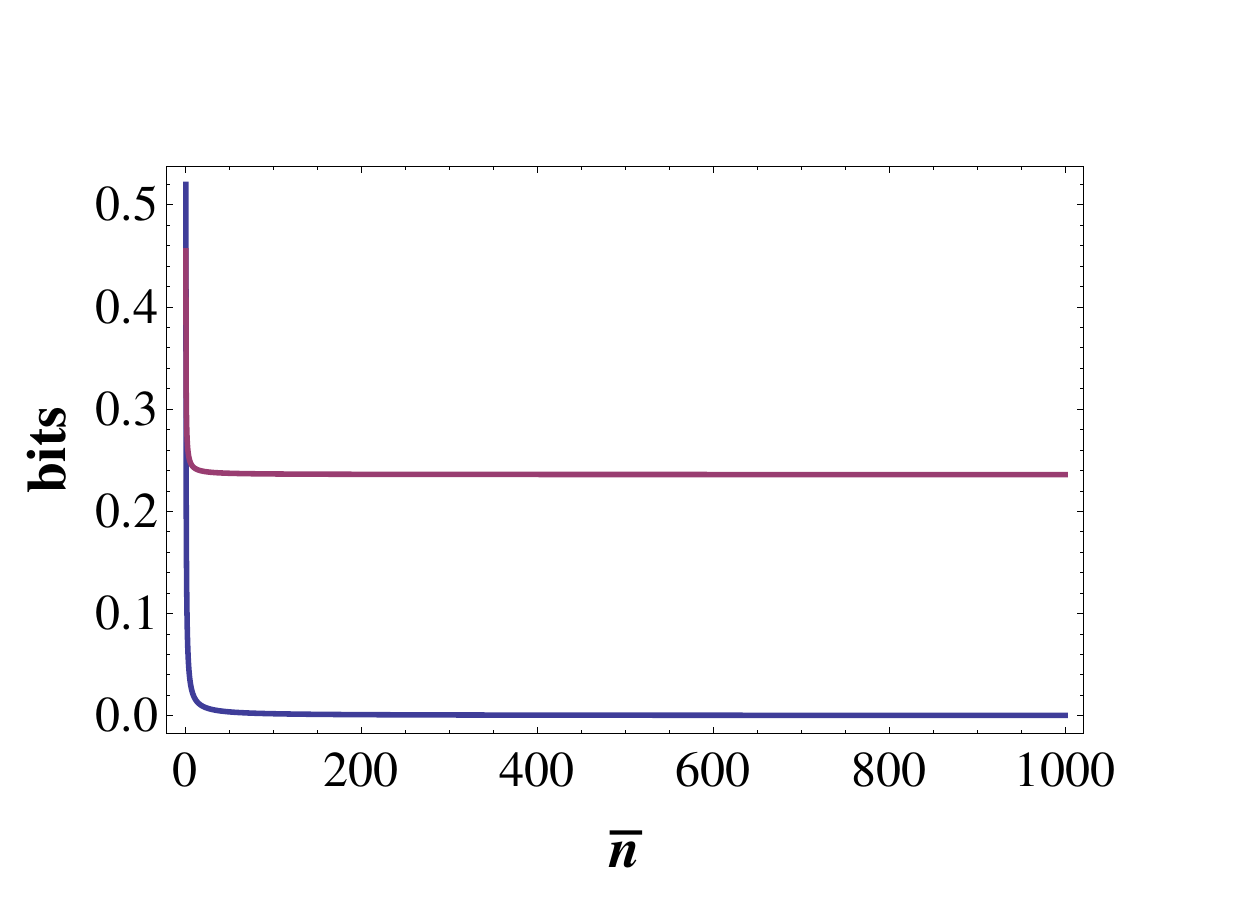} \vspace{-0.2cm}
\end{center}
\caption{We plot $I_{\text{read}}^{\text{class}}$ (lower curve) and $I_{%
\text{read}}^{\text{quant}}$ (upper curve)\ versus the mean photon number $%
\bar{n}\geq 1$. We assume a memory with reflectivity $r$ satisfying the
condition of Eq.~(\protect\ref{condR0}).}
\label{EPRpic2}
\end{figure}

\begin{figure}[tbph]
\vspace{-0.7cm}
\par
\begin{center}
\includegraphics[width=0.55\textwidth] {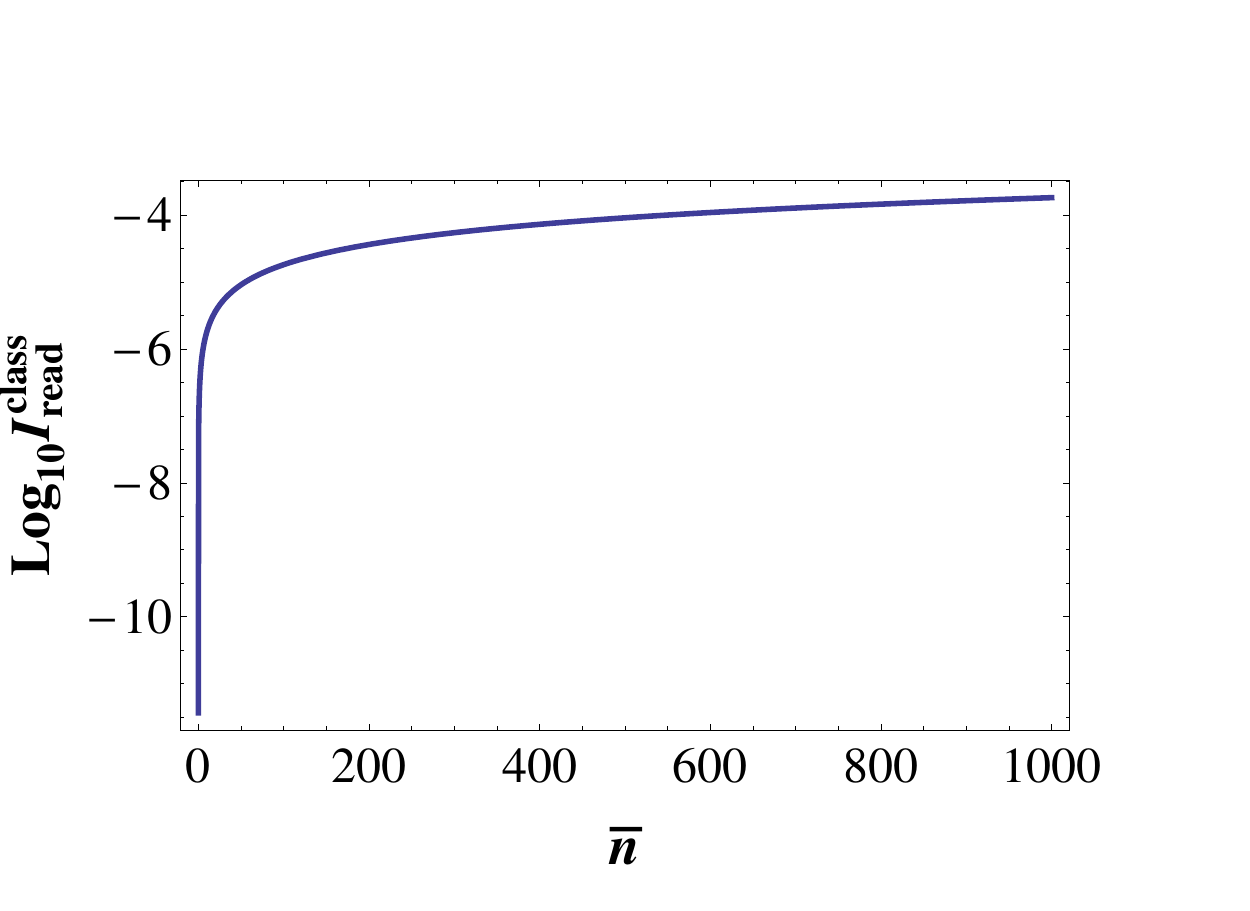}
\end{center}
\par
\vspace{-0.5cm} \caption{We plot the information quantity
$I_{\text{read}}^{\text{class}}$ in log-scale for $\bar{n}\leq
\bar{n}_{\text{max}}$. We consider the readout of a memory
with $1-r=\bar{n}_{\text{max}}^{-1}$. Here we consider the numerical value $%
\bar{n}_{\text{max}}=1000$ but the behaviour is generic.}
\label{EPRpic3}
\end{figure}

We can see that, at any fixed energy $\bar{n}$ irradiated over the memory
cell, there is a memory with reflectivity $r$ satisfying Eq.~(\ref{condR0})
which is readable by using a quantum transmitter\ with signal energy $\bar{n}
$\ but unreadable by a classical transmitter with the same irradiated energy
$\bar{n}$. More precisely, any classical transmitter with energy up to $\bar{%
n}$ is inefficient. In fact, let us fix some value $\bar{n}_{\text{max}}$
and consider a memory with $1-r=\bar{n}_{\text{max}}^{-1}$, then the
performance of all classical transmitters with signal energy $\bar{n}\leq
\bar{n}_{\text{max}}$ is shown in Fig.~\ref{EPRpic3}. We see that the
optimal classical transmitter is that with the maximal energy $\bar{n}_{%
\text{max}}$ as clearly expected from the monotonic expression in
Eq.~(\ref{CLASStransmitter}).

Thus, if we construct a\ theoretical memory which can be
irradiated with at most $\bar{n}_{\max }$ photons per cell
(otherwise data is lost, e.g., due to photodegrable effects) and
having reflectivity $r$ satisfying Eq.~(\ref{condR0}), then this
will be unreadable by any classical transmitter based on coherent
states while its data can be retrieved by a quantum transmitter
with signal energy $\simeq \bar{n}_{\max }$.

Note that in general, we can design a memory with reflectivity $r$ such that
\begin{equation}
1-r=c\bar{n}^{-1},  \label{condR}
\end{equation}%
for some constant $c$. For large $\bar{n}$, we have $I_{\text{read}}^{\text{%
class}}\rightarrow 0$, while $I_{\text{read}}^{\text{quant}}$ tends to a
constant $\leq 1$ which depends on $c$. For instance, we have $I_{\text{read}%
}^{\text{quant}}\rightarrow 0.895$ for $c=0.1$, and $I_{\text{read}}^{\text{%
quant}}\rightarrow 0.997$ for $c=0.01$. In the following Fig.~\ref{EPRpic4},
we show the behaviour of the two information quantities $I_{\text{read}}^{%
\text{class}}$ and $I_{\text{read}}^{\text{quant}}$ assuming the
condition of Eq.~(\ref{condR}) with $c=0.1$. We see how the
memories remains unreadable by classical means while the
peformance of quantum reading approaches 1 bit per cell.
\begin{figure}[tbph]
\vspace{-0.7cm}
\par
\begin{center}
\includegraphics[width=0.52\textwidth] {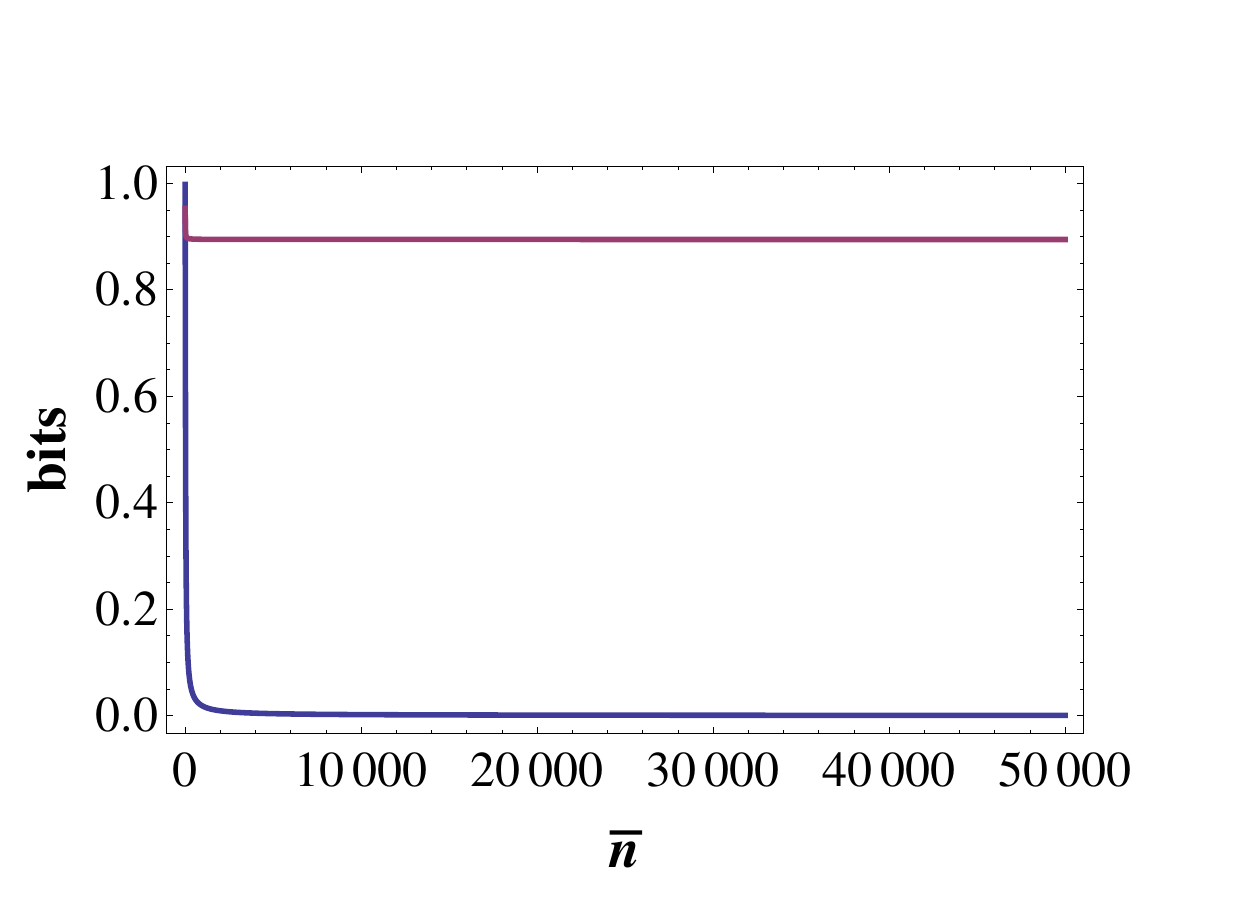} \vspace{-0.4cm}
\end{center}
\caption{We plot the information quantities $I_{\text{read}}^{\text{class}}$
(lower curve) and $I_{\text{read}}^{\text{quant}}$ (upper curve)\ versus the
mean photon number $\bar{n}\geq 1$. We consider memories with reflectivity $%
r $ satisfying Eq.~(\protect\ref{condR}) with $c=0.1$.}
\label{EPRpic4}
\end{figure}

\section{Conclusion}

\label{sec3}

In this preliminary study on the cryptographic aspects of quantum
reading, we have shown how it is possible to construct classical
memories which cannot be read by classical means, namely coherent
states (and mixtures of coherent states, by invoking the same
convexity arguments of Ref.~\cite{pirandola}) but still they can
be read using quantum entanglement. In particular, we have
considered an EPR state and we have connected the mean number of
photons to be employed by this quantum source with the
reflectivies to be used in the memory cells, see
Eq.~(\ref{condR0}) and also its generalization in
Eq.~(\ref{condR}). Note that other non-classical states may also
provide non-trivial advantages with respect to coherent states and
their mixtures. In general, the security provided by the scheme
relies on the technological difference between two types of labs,
one limited to classical sources and the other able to access
quantum features, such as entanglement or squeezing.

It is interesting to discuss the connections between our scheme of
data-hiding by quantum reading and the traditional technique of
quantum data hiding~\cite{QDhide1,QDhide2}. The latter is about to
store classical information into entangled states, so that it can
only be retrieved by joint measurements. It is clearly an
application of quantum state discrimination. By contrast,
data-hiding by quantum reading is related to the problem of
quantum channel discrimination. Classical data is stored in a
channel (not a state) and quantum entanglement is used as an input
resource to be processed by the channel. This is a crucial
difference, also for practical purposes, since data stored in a
classical memory does not decohere (like the entangled states
typically prepared in quantum data hiding), and quantum
entanglement is used a resource on demand, which is needed only
for the readout of the information (not for the storage process).

Note that our study can be extended in several ways. We have only
considered ideal memories where the cells are addressed
individually and have very high reflectivities (in particular, we
have assumed unit reflectivity for one of the two bit values
stored in the cell). There is no inclusion of additional noise
sources in the model, e.g., coming from stray photons scattered
during the readout process, neither analysis of diffraction or
other optical effects. Finally, we have also assumed that high
values of entanglement can be generated. While this is possible
theoretically, it is very hard to achieve experimentally. This
would not be a problem if we were able to construct memories which
are extremely photo-sensitive, so that that the maximum values of
tolerable energies are of the order of $\bar{n}_{\max }\lesssim
10$ photons per cell.

\section*{Acknowledgments}

This work is supported by an EPSRC DTG grant (UK). G.S. would like
to thank C. Ottaviani, S. L. Braunstein, S. Mancini and  S.
Pirandola for useful discussions.

\end{document}